# Algorithms for Improving the Automatically Synthesized Instruction Set of an Extensible Processor

P.N. Sovietov, RTU MIREA, sovetov@mirea.ru

Processors with extensible instruction sets are often used today as programmable hardware accelerators for various domains. When extending RISC-V and other similar extensible processor architectures, the task of designing specialized instructions arises. This task can be solved automatically by using instruction synthesis algorithms. In this paper, we consider algorithms that can be used in addition to the known approaches and improve the synthesized instruction sets by recomputing common operations (the result of which is consumed by multiple operations) of a program inside clustered synthesized instructions (common operations clustering algorithm), and by identifying redundant (which have equivalents among the other instructions) synthesized instructions (subsuming functions algorithm).

Experimental evaluations of the developed algorithms are presented for the tests from the domains of cryptography and three-dimensional graphics. For Magma cipher test, the common operations clustering algorithm allows reducing the size of the compiled code by 9%, and the subsuming functions algorithm allows reducing the synthesized instruction set extension size by 2 times. For AES cipher test, the common operations clustering algorithm allows reducing the size of the compiled code by 10%, and the subsuming functions algorithm allows reducing the synthesized instruction set extension size by 2.5 times. Finally, for the instruction set extension from Volume Ray-Casting test, the additional use of subsuming functions algorithm allows reducing problem-specific instruction extension set size from 5 to only 2 instructions without losing its functionality.

**Introduction**

Programmable hardware accelerators, specifically designed for limited classes of tasks, are becoming more widely used. Such accelerators, by specializing their hardware, can offer better performance, in particular energy efficiency, than general-purpose processors. The diversity of today's programmable hardware accelerators, aimed at such classes of tasks as, for example, cryptography, image processing or machine learning, allows to say that a "new golden age of computer architecture" has arrived [1].

One of the approaches to designing the architecture of a programmable hardware accelerator is specialization of the general-purpose processor data path with expansion of the instruction set with domain-specific instructions. In many cases, the designer's task is to introduce the minimum number of new instructions that provide the required speedup on the target class algorithms. There are a number of processor architectures with extensible instruction sets (extensible processors): Tensilica Xtensa, ARC, Arm Custom Instructions, RISC-V. There are also a number of CAD systems for extensible processors in which the design is performed using domain-specific instruction set description languages. Examples of such languages are TIE (Xtensa processor generator), nML (ASIP Designer), and CodAL (Codasip Studio). It is assumed that the designer creates the description of the required instruction set manually, i.e. the task of automatic synthesis of specialized instructions is not solved in these CAD systems.

Automatic synthesis of specialized instructions for an extensible processor can be based on code analysis of algorithms of the target class of tasks. The goal is to extract computation subgraphs: the most promising areas of hardware acceleration at different levels of program structure and depending on the given architectural constraints. Such typical subgraphs are candidates for specialized instructions for an extensible processor. There is a number of algorithms for synthesis of specialized instructions [2] based on the analysis of the basic blocks of the program.

This paper considers algorithms that can be used in addition to approaches known from the literature and that achieve an additional improvement in the set of synthesized instructions of an extensible processor. This additional improvement is achieved by recomputing common operations in the extended synthesized instructions (common operations clustering algorithm), as well as by identifying redundant synthesized instructions that are special cases of computations of other instructions (subsuming functions algorithm). The experimental evaluations presented in this paper, based on the analysis of program code from the field of cryptography as well as from the field of 3D graphics, demonstrate the reduction in the size of compiled programs and the reduction in the total number of synthesized instructions obtained using the developed algorithms.

**Synthesis of the instruction set with its subsequent improvement**

The instruction synthesis process consists, in general, of the following steps::

1. Extraction of instruction subgraphs from the data dependency graph (DDG) of the program according to the specified architectural constraints.
2. Filtering the set of extracted subgraphs in order to get rid of functionally equivalent duplicates and obtain a minimal set of the most preferred candidates in terms of hardware acceleration potential.

Synthesized instructions have the following common forms:

- single-output subgraph (Multiple-Input Single-Output, MISO);
- subgraph with multiple outputs (Multiple-Input Multiple-Output, MIMO);
- SIMD-MIMO – an unconnected MIMO variant focused on SIMD computing.

The most efficient instruction synthesis algorithms are designed to extract MISO subgraphs on the basic block level. One of the simplest among such algorithms is the greedy MaxMISO algorithm [3].

Figure 1 shows the DDG of the Magma cipher round. The gray color indicates the subgraphs-clusters found using MaxMISO. It can be seen that the addition operation marked in blue color did not fall into any of the found clusters, since it is a common operation.

A common operation in DDG is an operation whose outgoing edges go to the inputs of at least two other operations. In other words, the result of the computation of a common operation is used as arguments by two or more other operations.

If it were possible to get rid of the common addition operation shown in Fig. 1, it would not only reduce the size of the compiled code, but also make it possible to get rid of an extra cycle of calculations using instruction parallelism. This effect can be achieved by copying common operations and recomputing them.

Fig. 2 shows the results of augmenting MaxMISO with the output of the common operations clustering algorithm discussed in the next section. It can be seen that the clusters of synthesized instructions are now enlarged and the size of the compiled encryption round code for the RISC architecture is reduced by one instruction. This algorithm can be used not only with MaxMISO, but also with other well-known MISO subgraph extraction algorithms.

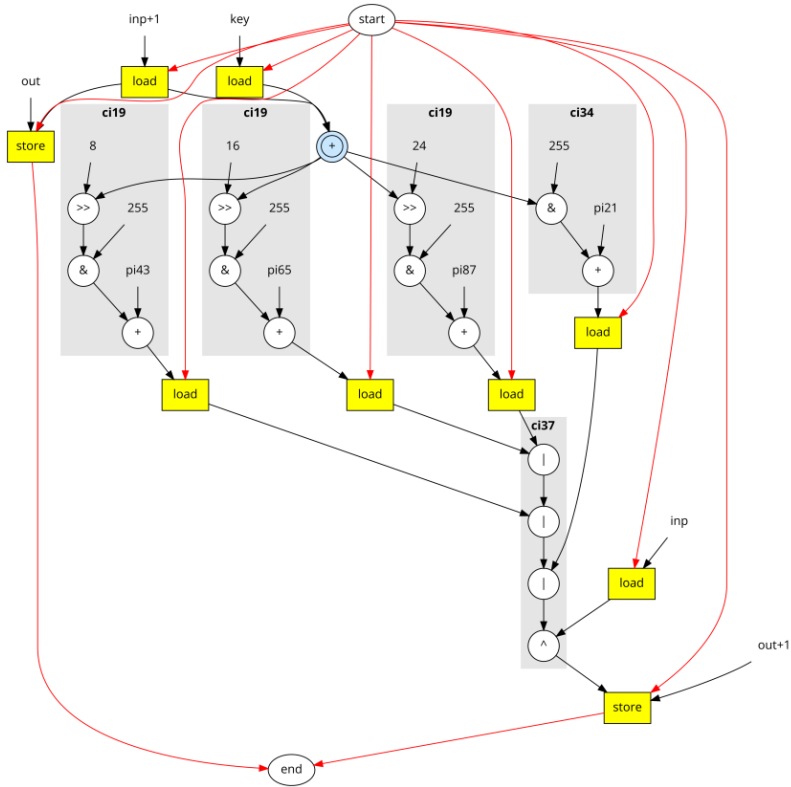

*Fig. 1. Results of the MaxMISO algorithm execution for the Magma cipher round. The gray color indicates the found subgraphs, and the blue color indicates the common operation*

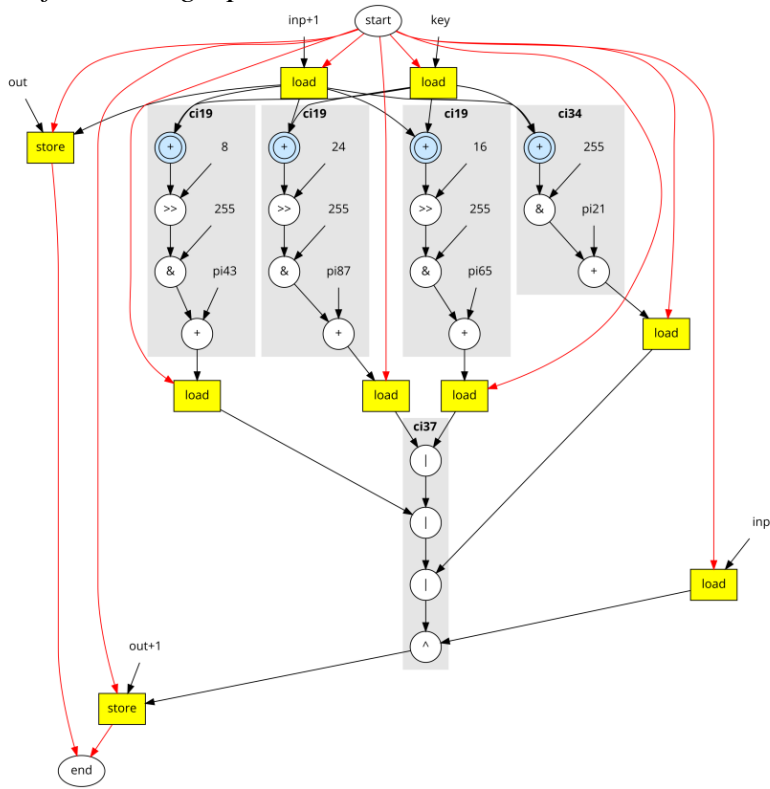

*Fig. 2. MaxMISO algorithm execution results for the Magma cipher round with the addition of the common operations clustering algorithm*

From Fig. 2, it can be concluded that speeding up the Magma encryption round requires adding three new synthesized instructions to the instruction set of the extensible processor: ci19, ci34, and ci37. Instruction ci34 is a special case of ci19 computation:

```
ci19(a, b, c, d) := (((a + b) >> c) & 255) + d
   ci34(a, b, c) := ((a + b) & 255) + c
```

Thus, to speed up a round of Magma encryption, it is sufficient to have two synthesized instructions if it is possible to set constant arguments for ci19:

```
ci19(a, b, 0, c) = ci34(a, b, c)
```

To identify redundant synthesized instructions that are special cases of other instructions, the subsuming function algorithm described in this paper is developed.

**Common operations clustering algorithm**

The main idea of the iterative algorithm for clustering common operations is that the coverage of the program's DDG by subgraphs can be more complete if we include in these subgraphs, as long as it is possible, copies of the common operation with its recomputation.

The considered algorithm is shown in Fig. 3. Its main stages are:

– single node clustering (lines 21–22);
– traversal of clusters in topological sorting order (lines 10–17);
– merging a MISO-cluster with its "users" (line 14);
– repetition as long as there are changes (lines 23–26).

The *is_legal_miso* function defined in lines 1–2 checks the "synthesizability" of the MISO-subgraph: the subgraph must be convex and satisfy a set of architectural constraints.

This algorithm is used in the final stage of instruction subgraph extraction, after applying one of the known MISO subgraph extraction algorithms, The proposed algorithm can also be used as a single instruction subgraph extraction algorithm, but in general with slightly worse DDG coverage performance compared to MaxMISO.

```
1:  function IS_LEGAL_MISO(G, S)
2:      return (∀i ∈ S \ root(S). ∀u ∈ uses(G[i]). u ∈ S) ∧ arch_constraints(G, S)
3:  function COMBINE_SUBGRAPHS(G, u, i)
4:      return cluster((inputs(G[u])∪inputs(G[i]))\{i}, nodes(G[u])∪nodes(G[i]))
5:  function CAN_COMBINE(G, u, i)
6:      return is_legal_miso(G, nodes(G[u]) ∪ nodes(G[i]))
7:  function COMBINE(G)
8:      G' ← ∅
9:      M ← ∅
10:     for each i ∈ toposort(G, i_{end}) do
11:         if i ∉ M then
12:             if ∀u ∈ uses(i). can_combine(G, u, i) then
13:                 for each u ∈ uses(i) do
14:                     G'[u] ← combine_subgraphs(G, u, i)
15:                     M[u] ← 1
16:             else
17:                 G'[i] ← G[i]
18:     return G'
19: function CLONE_AND_COMBINE(G)
20:     G' ← ∅
21:     for each i ∈ G do
22:         G'[i] ← cluster(inputs(G[i]), nodes(G[i]))
23:     repeat
24:         G ← G'
25:         G' ← combine(G)
26:     until G = G'
27:     return G
```

*Fig. 3. Common operations clustering algorithm*

**Subsuming functions algorithm**

A function *f* subsumes a function *g* if a tuple of arguments of *f* can be composed of arguments of *g* and constants from *c*, such that *f* and *g* become input-output equivalent:

$$\exists p_1 \in [1..n] \ldots \exists p_n \in [1..n]. \left(\forall x_1 \ldots \forall x_m. f(v_{p_1}, \ldots, v_{p_n}) = g(x_1, \ldots, x_m)\right),$$
$$v = (x_1, \ldots, x_m, c_1, \ldots, c_{n-m}), n \geq m.$$

Thus, the problem of determining whether a function *f* subsumes a function *g* can be solved by choosing appropriate names of arguments and values of constants. This problem is reduced to the SMT (Satisfiability Modulo Theories) problem and an SMT solver can be used to solve it.

Fig. 4 shows the subsuming functions algorithm. This algorithm uses the Counter-Example Guided Inductive Synthesis (CEGIS) technique [4].

The main elements of the algorithm considered are:

– *subsume* function implementing CEGIS loop of synthesis and verification of the next synthesized solution (lines 21–28);
– *synth* function for synthesizing operands taking into account the set of tests generated with the *verify* function (lines 1–11).

The synthesized model contains the selected indices *p* of the argument names of the function *f*, as well as the values of the constants *c*.

```
1:  function SYNTH(f, g, T)
2:      p ← (p_1 ∈ [1..n], ..., p_n ∈ [1..n])
3:      c ← (c_1, ..., c_{n-m})
4:      for each (x, y) ∈ T do
5:          v ← (v_1, ..., v_n)
6:          SMT.add(⋀_{i=1}^{n} (⋁_{j=1}^{m} (p_i = j ∧ v_i = x_j) ∨ ⋁_{j=m+1}^{n-m} (p_i = j ∧ v_i = c_{j-m})))
7:          SMT.add(⋀_{i=1}^{m} ⋁_{j=1}^{n} x_i = v_j)
8:          SMT.add(f(v_1, ..., v_n) = g(x_1, ..., x_m))
9:      if SMT.check() = sat then
10:         return SMT.model(), true
11:     return ⊥, false

12: function VERIFY(f, g, m)
13:     x ← (x_1, ..., x_m)
14:     v ← get_v(m.p, x, m.c)
15:     SMT.add(y = f(v_1, ..., v_n) ∧ y ≠ g(x_1, ..., x_m))
16:     if SMT.check() = sat then
17:         return (x, y), false
18:     return ⊥, true
19: function SUBSUME(f, g)
20:     T ← ∅
21:     loop
22:         m, err ← synth(f, g, T)
23:         if err then
24:             return ⊥, false
25:         t, err ← verify(f, g, m)
26:         if ¬err then
27:             return m, true
28:         T ← T ∪ {t}
```

*Fig. 4. Subsuming functions algorithm based on operands synthesis*

**Experimental evaluations of the developed algorithms**

The developed algorithms were tested on software implementations of Magma and AES block ciphers. Using the methods outlined in [5], variants of the retargetable compiler with code generation for the program model of the processor having RISC-V RV32I architecture extended with synthesized instructions were implemented. MaxMISO with architectural constraints on the maximum number of inputs of the extracted subgraph in the range of 1–6 was used as a basic algorithm for extracting subgraphs of synthesized instructions.

Fig. 5 shows the compiled code size data for the software implementation of the Magma cipher using various variants of the synthesized instructions extension.

In the Magma test, clustering of common operations reduces the code size starting from the variant with synthesized instructions constrained by 3 inputs. The maximum code size reduction is observed on the variant with 6 inputs constraint and is 9%.

Fig. 6 shows similar data for the AES cipher. Clustering of common operations reduces the code size starting from the variant using instructions also constrained by 3 inputs. The maximum code size reduction is obtained for the variant constrained to 6 inputs and is 10%.

The data on the application of the subsuming functions algorithm for Magma and AES ciphers are summarized in Table 1. The maximum reduction of the set of synthesized instructions for Magma cipher is 2 times, and for AES – 2.5 times.

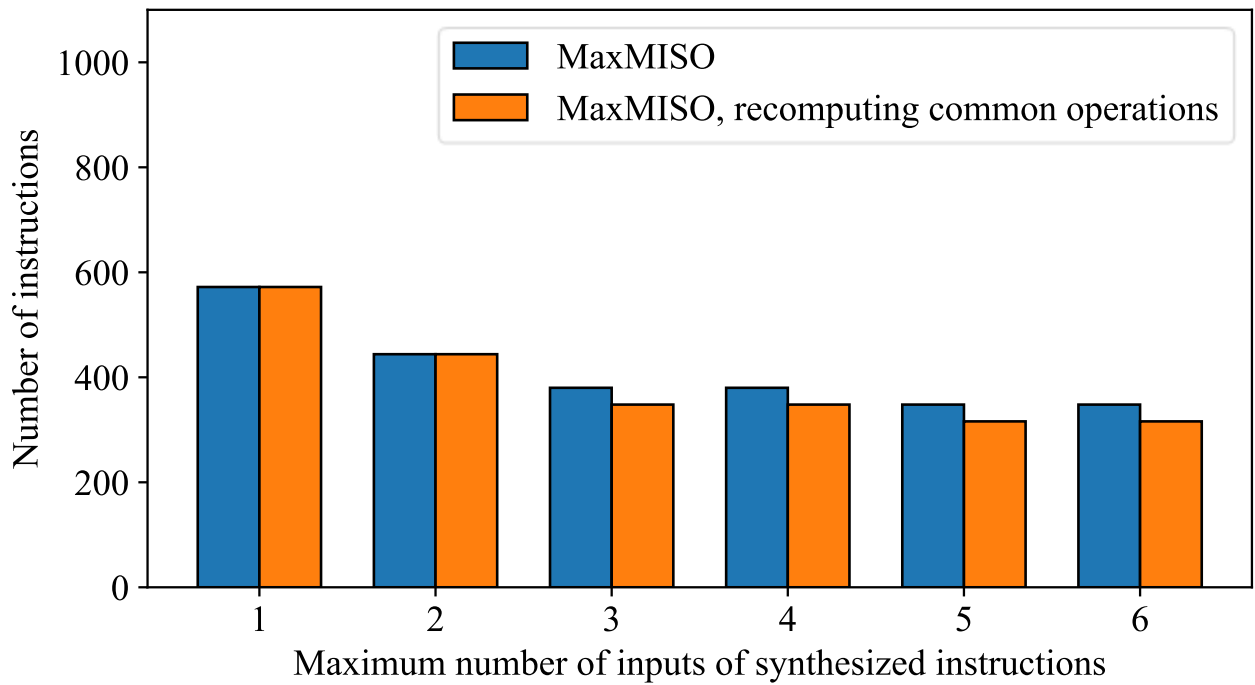

*Fig. 5. Application of the common operations clustering algorithm for the implementation of the Magma cipher*

Table 1. Results of the subsuming functions algorithm execution for the implementations of Magma and AES ciphers

| Maximum number of inputs | Magma (reducing of the synthesized instruction set, times) | AES (reducing of the synthesized instruction set, times) |
|---|---|---|
| 2 | 1.7 | 2.5 |
| 3 | 1.7 | 1.2 |
| 4 | 1.7 | 2.3 |
| 5 | 2 | 1.2 |
| 6 | 2 | 1.3 |

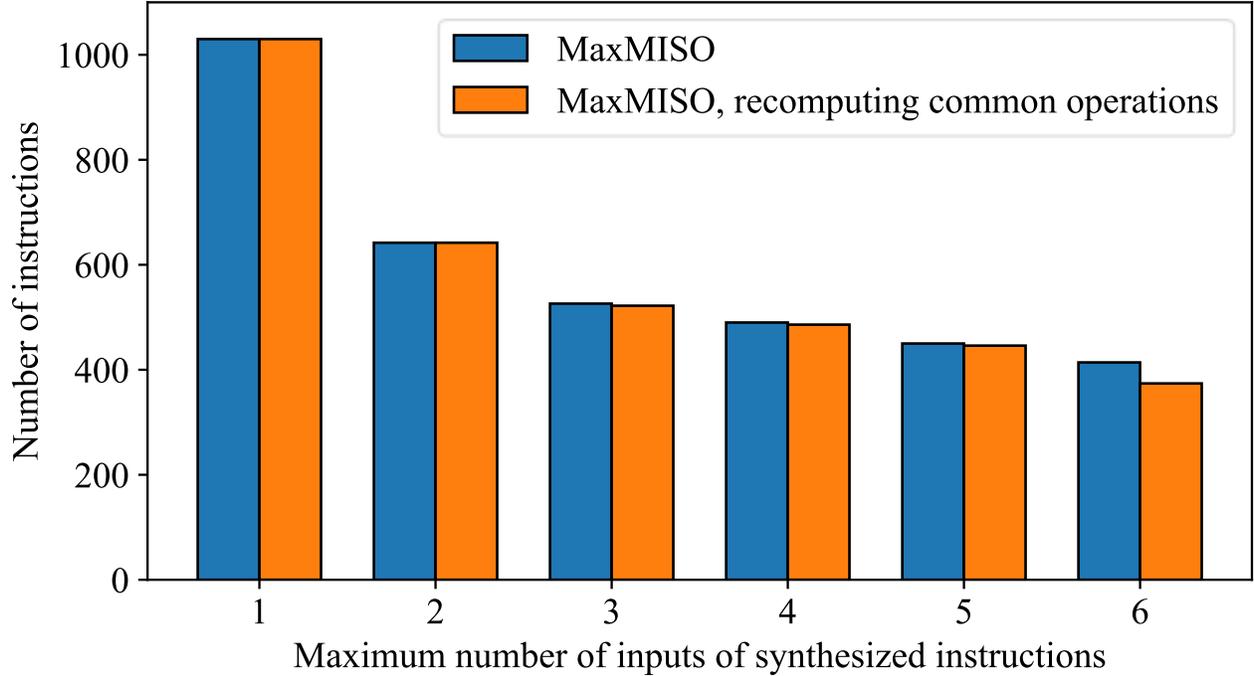

*Fig. 6. Application of the common operations clustering algorithm for the implementation of the AES cipher*

In [6], examples of the results of the MISO-instruction synthesis algorithm are presented. In particular, this work shows an extension of 5 synthesized instructions extracted from the Volume Ray-Casting software test for 3D graphics. With the additional use of the subsuming functions algorithm, the number of synthesized commands is reduced to only 2 instructions, as can be seen from Table 2.

Table 2. Results of the subsuming functions algorithm execution for the Volume Ray-Casting test from [6]

| Synthesized instruction | Subsuming result |
|---|---|
| fmadd(A, B, C) := ((A * B) + C) | fdot(1, C, B, A, 0, 0) |
| f2madd(A, B, C, D) := ((A * B) + (C * D)) | fdot(A, B, D, C, 0, 0) |
| f3madd(A, B, C, D, E) := (((A * B) * C) * D) + E | ⸺ |
| fdot(A, B, C, D, E, F) := ((A * B) + (C * D)) + (E * F) | ⸺ |
| fmmul(A, B, C) := ((A * B) * C) | f3madd(C, 1, B, A, 0) |

**Overview of known instruction synthesis algorithms**

Table 3 summarizes the characteristics of some known instruction synthesis algorithms operating at the basic block level. Using these algorithms, convex subgraphs are extracted from the DDG of the program, without common operations for several subgraphs. In [7], common operations are considered only in special conditions: at the stage of instruction selection, to correct the results of the algorithm for generating subgraphs with overlaps.

At the filtering stage, known instruction synthesis algorithms do not take into account functional equivalence of subgraphs and do not identify cases when one subgraph is a special case of another subgraph. In [8] and [9], heuristics applicable only to special cases are used, and the data path merging technique [10] requires additional multiplexer synthesis.

Table 3. Algorithms for instruction set synthesis on the basic block level

| Reference | Subgraph form | Constraints | Extraction | Filtering |
|---|---|---|---|---|
| Alippi [3] | MISO | Maximal size | Greedy incremental clustering algorithm (MaxMISO) | No |
| Atasu [11] | Arbitrary | Architectural | Incremental clustering based on the ILP problem | Isomorphism-based |
| Pozzi [12] | Arbitrary | Architectural | Optimal graph partitioning | No |
| Pulli [13] | connected MIMO | No | Finding the maximum common subgraph | Isomorphism, graph covering problem |
| Xiao [14] | connected MIMO | No | Enumeration of all subgraphs | No |

**Conclusion**

The paper considers the task of instruction set synthesis of an extensible processor and gives an overview of known instruction set synthesis algorithms. It is shown that the improvement of these algorithms can be achieved by recomputing common operations in the larger synthesized instructions, as well as by identifying redundant synthesized instructions, which are special cases of computing of other instructions. The developed algorithms that additionally improve the performance of known MISO-graph extraction algorithms are presented: the common operations clustering algorithm and the subsuming functions algorithm.

Experimental evaluations of the developed algorithms are obtained, demonstrating, in particular, that for the software implementation of the AES cipher the common operations clustering algorithm allows to reduce the size of the compiled code by 10%, and the subsuming functions algorithm for the same AES cipher implementation reduces the set of synthesized instructions by 2.5 times.